\documentclass[runningheads,a4paper]{iaga}

\usepackage{amssymb}
\usepackage{graphicx}

\usepackage{url}
\newcommand{\keywords}[1]{\par\addvspace\baselineskip
\noindent\keywordname\enspace\ignorespaces#1}

\begin{document}

\mainmatter  

\title{Ground-based measurements of solar diameter}

\titlerunning{Overview on Solar Diameter Sigismondi}

\author{Costantino Sigismondi}

\authorrunning{C. Sigismondi}

\institute{ICRA-Sapienza, P.le Aldo Moro 5 00185 Rome Italy;\\
 University of Nice-Sophia Antipolis, Dept. Fizeau UMR 6525 Nice France;\\
 IRSOL, via Patocchi, 6610 Locarno Monti, Switzerland\\
\email{sigismondi@icra.it}
\\ home page:
\texttt{http://www.icra.it/solar}}


\maketitle

\begin{abstract}
The solar diameter changes or not? Whatever will be the answer the methods used for its measurements are more and more challenging, and facing new astrophysical and optical problems since the required space resolution is of astrometric quality.
A quick overview on different methods is here presented, as well as the problem of the solar limb definition, emerging after the flash spectrum during eclipses.
\end{abstract}

\section{Introduction}
The measurements of the solar diameter have been made sistematically in the XIX century. At the end of that century Auwers\cite{Auwers} published a value for the solar radius of 959.63 arcseconds. This value is now adopted as standard value from IAU, the International Astronomical Union.
Frahunhofer invented at the beginning of XIX century the heliometer. It is the prototype of direct measurement of the angular diameter of the Sun. This instrument was so accurate that allowed F. W. Bessel to measure in 1838 the first parallax of a star: 61 Cygni, 0.3 arcsec, selected for its significant proper motion of 5.2 arcsec/yr already discovered in 1812 by G. Piazzi. The heliometer in Goettingen\cite{Goettingen} (1895) was a conceptual advancement of the heliometer's design; its space version is the Solar Disk Sextant SDS and its educational one is the double pinhole solar monitor\cite{Sigi02,Sigi06}.
The measurements of the solar diameter by meridian transit were monitored on a daily basis since 1851 at Greenwich Observatory and at the Campidoglio (Capitol) Observatory in Rome since 1877 to 1937\cite{Gething}. Afterwards solar astrolabes (Danjon type and DORAYSOL\cite{Doraysol}, Definition et Observation du RAYon SOLaire, have obtained similar results with lower scatters. These methods have in common the use of a fixed telescope and the observation of the drift of the solar image through a meridian or a given {\it almucantarat}. 
\footnote{Almucantarat is an arabic name standing for circle of the same altitude above the horizon.} 
Eclipses and planetary transits exploited the timing of the orbital motion of the Earth, Moon and Planets; their angular velocity is much slower than the daily motion of the Sun (geocentric view) and they can allow very accurate timing determination. Black drop and seeing effects can be overcome for planetary transits by fitting the chord draft by the planet's disk over the solar limb with an analitycal function\cite{SolObs}. With fast video recording either eclipses or drift-scan transits can also achieve interesting timing resolution. Eclipses data of solar diameter still show a random scatter of 0.5 arsceconds\cite{Sigi09} around the standard value. 
Planetary transits and eclipses are space measurements in the table of fig. 1, even if these observations are ground-based, because the influences of seeing are limited, e.g. the dis/appearance of a bead and this is an on/off signal and seeing acts only through scintillation\cite{SigiNugentDangl}. RHESSI\cite{Fivian} measurements of the solar oblateness (of general relativistic interest\cite{Sigi05,Sigi05b}) are better than previous ground-based measurements\cite{Dicke}. In recent publications of SOHO satellite group\cite{Kuhn04,Kuhn10} data are interpreted as if the Sun has a rock-steady diameter.  
\begin{figure}
\centering
\includegraphics[width=12cm]{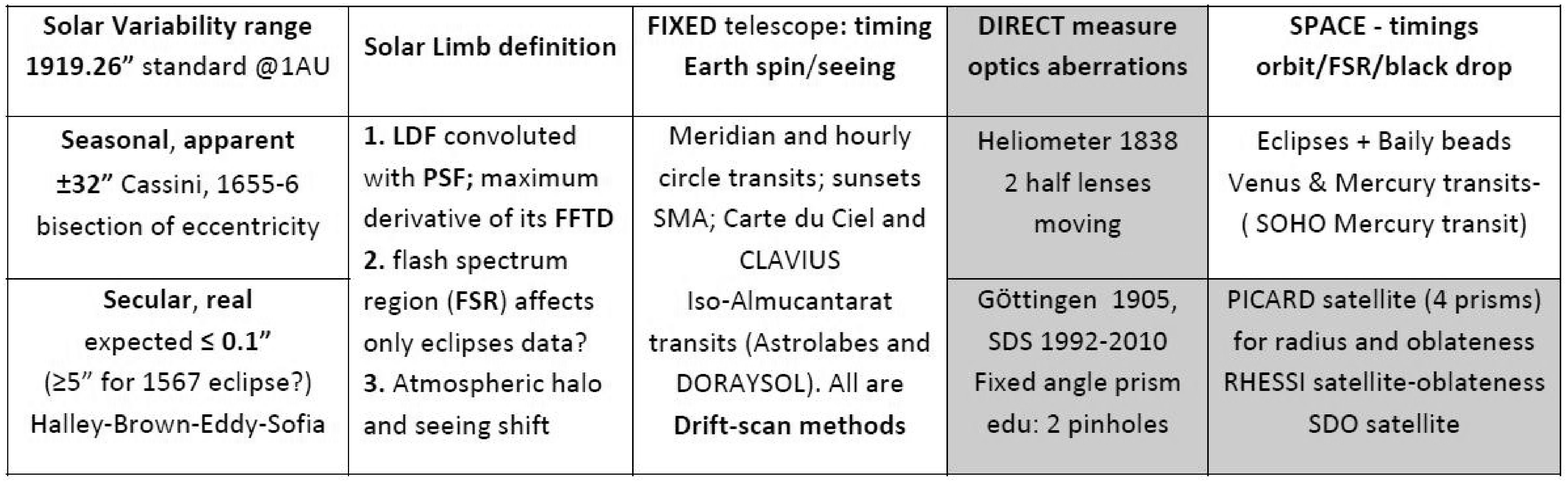}
\caption{In this table are indicated the ranges of variability of solar diameter and the highlights on solar limb definition. The three methods for diameter's measurement are classified with respect to their reference angular velocity and their limiting factors (atmospheric seeing, black-drop effect or Flash Spectrum Region): daily rotation of the Earth rules drift-scan methods (FIXED); monthly orbit of the Moon and eclipses and yearly orbits of planetary transits give the typical velocities of the other methods (SPACE timings); the DIRECT measurements are limited by seeing and by optical aberrations of instruments, excepted in the case of pinholes. Among direct measurements there are also space satellite: the shadowed cells are all belonging to direct measurements. FSR Flash Spectrum Region is discussed in paragraph 3.}
\end{figure}

\begin{figure}
\centering
\includegraphics[width=12cm]{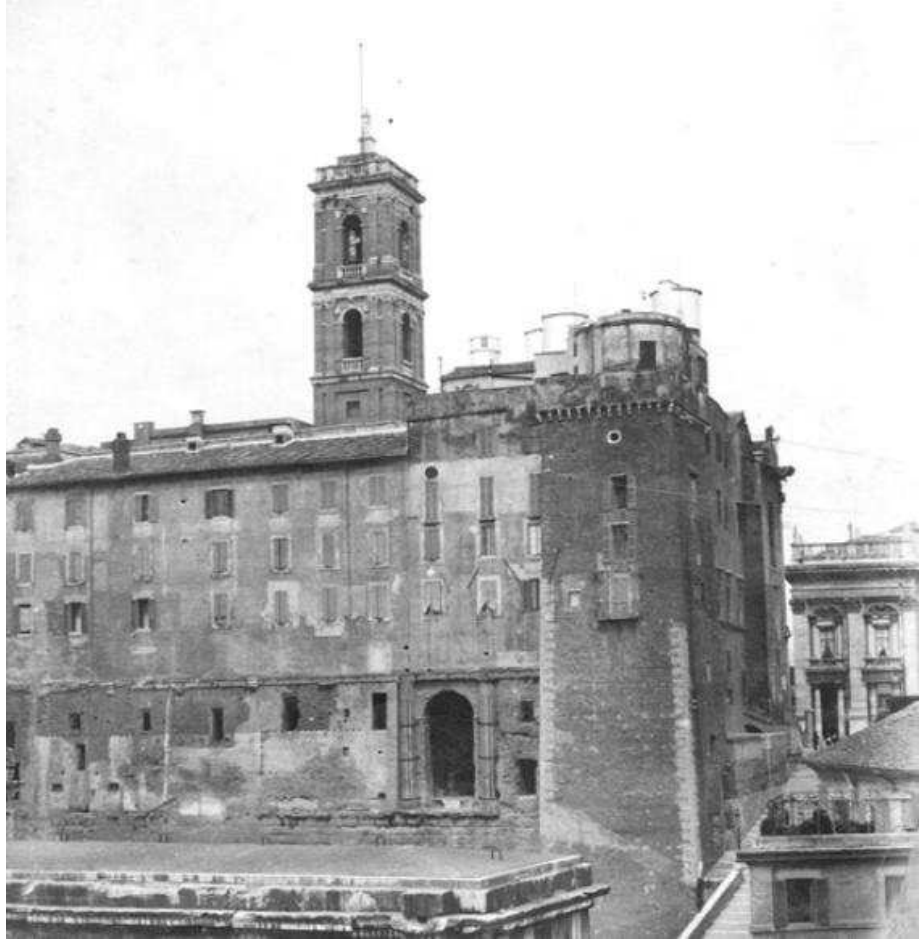}
\caption{The observatory of Campidoglio in Rome, as before 1937 when it was destroyed. It has been founded by pope Leo XII in 1829. Ancient photo, courtesy of Renzo Lay.}
\end{figure}

\begin{figure}
\centering
\includegraphics[width=12cm]{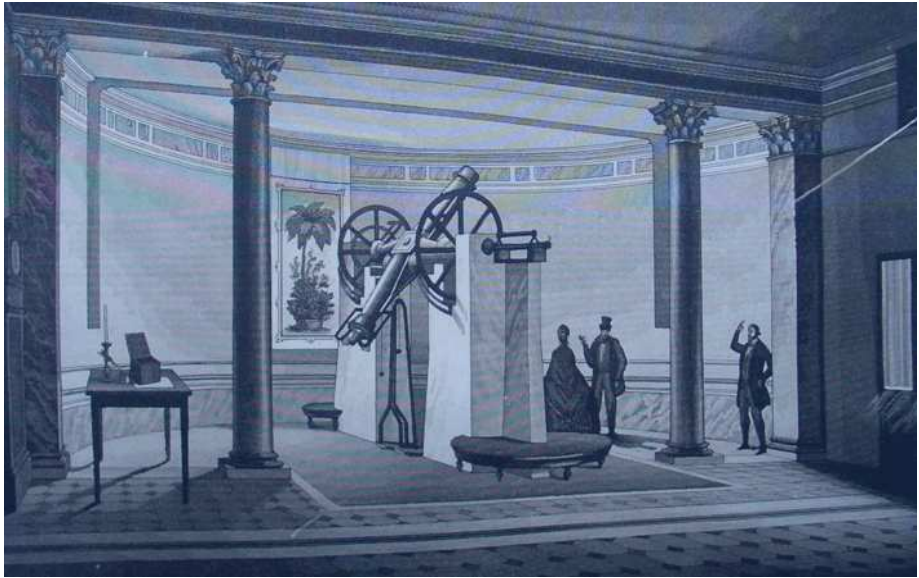}
\caption{The meridian circle in the observatory of Campidoglio in Rome, where all the measurements of solar diameter at the meridian transit were carried out. Ancient print, courtesy of Renzo Lay.}
\end{figure}

\begin{figure}
\centering
\includegraphics[width=12cm]{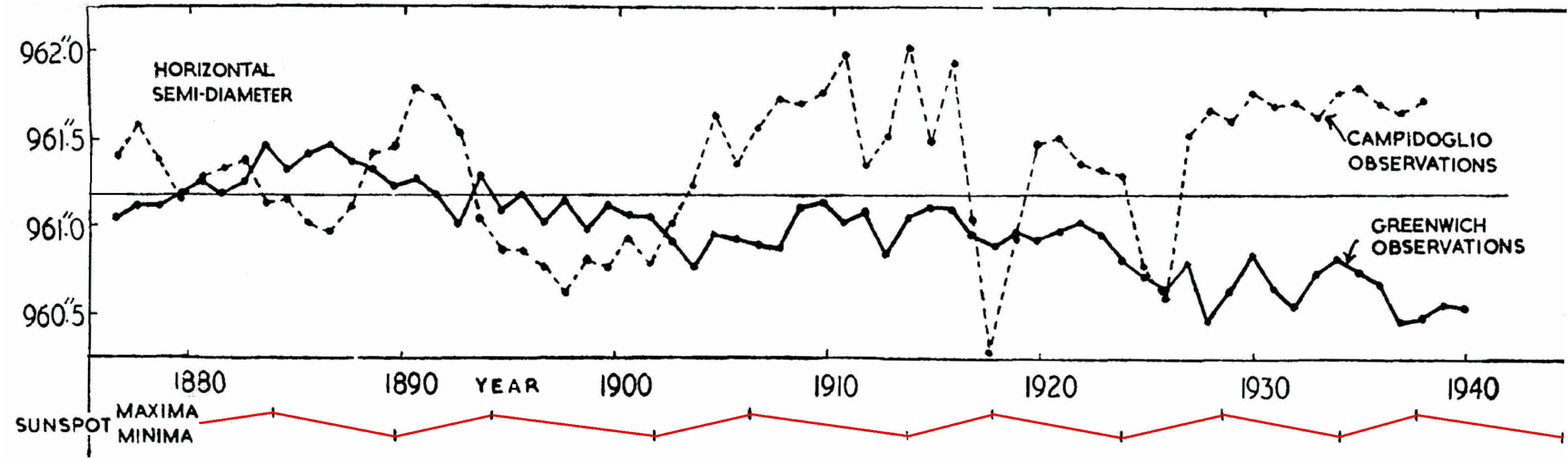}
\caption{The values of the solar radius measured at Campidoglio and at Greenwich compared (adapted from Gething, 1955). The difference between the two measurements is due to the atmospheric effects, as well as, to personal equations of the observers with different sensitivity to wavelength and timing. The measurements, all made with naked eye observations, were expected to show small errorbars, thanks to the statistical average. Each point is the annual average of the measurements and the statistical convergence is surprisingly missing. The results of this method seem to be affected by some systematical effects randomly different from one year to another... a contradictory affirmation which shows the troubles that all drift-scan methods have done up to DORAYSOL experiment (1999-2008).}
\end{figure}

\section{The eclipses and the Baily's beads}
The debate on the long-term variability of solar diameter started in 1978\cite{Eddy} when ancient eclipse data where used to demonstrate a variation in the solar diameter in the past 300 years. The duration of totality is a rapid function ($\Delta T\sim \sqrt d$) of the distance $d$ between the observer and the limit of the umbra, consequently near the borders of the totality path, where the eclipse is nearly grazing (the North or the South pole of the Moon moves nearly tangent to the photosphere)\cite{Sigi08}, it is possible to know the distance of the observer from the actual limit of umbra. 
Thanks to the rapid variation of the duration $\Delta T$ of the umbral phase, an observation of a few seconds (one or two) of totality and the unambiguous identification of the position of the observer, are parameters sufficient to know precisely, $\pm$ 10 m, this distance, and a great relative precision can be achieved with respect to the whole extension of the umbra on the Earth's surface $\sim$ 300 Km. This is the same relative accuracy achieved on the solar diameter determination with central eclipses.  
If there are observers located at both shadow's limits, North and South, the uncertainty on the ephemerides adopted in the data analysis can be also  bypassed. The last uncertainties arise from the adopted lunar profiles, such as Watts\cite{Watts} or Kaguya\cite{kaguya}, and from lunoid's corrections, such as Morrison and Appleby\cite{MA} or Soma\cite{soma}.
From the total eclipse observed by Halley in 1715 two observers located near the two borders were identified and an hypothesis on the past value of solar radius was made, namely 0.48 arcseconds larger than its standard value of 959.63 arcseconds.
This extra-radius should have been reducing over the following 200 years, up to 1925, when a total eclipse was studied in Yale University under prof. Brown's guidance. Southern limit was on Manhattan Island (New York City), while Northern limit was identified near Ladd Observatory (Providence, Rhode Island) where a flash spectrum
\footnote{The flash spectrum is an array of emission lines detectable from the limb of the Sun during the flash periods of a few seconds just after the beginning of totality during a solar eclipse or just before the instant of its termination. When the solar photosphere is occulted by the Moon, the layers of the Sun's atmosphere flash into prominence, and the spectrum briefly shows the bright lines at all wavelengths produced by tenuous hot luminous gas. Except during eclipses, this part of the spectrum is masked by the glare of the Sun's disk. Study of the flash spectrum gives information about the physical state of the solar chromosphere. The flash spectrum was first observed by the American astronomer Charles Augustus Young during the eclipse of Dec. 22, 1870.} 
was observed with an objective prism\cite{Paddock}. In 1979 after 3 whole Saros cycles, another eclipse casted its shadow on USA and the analysis confirmed that the radius was similar to 1925.
David Dunham proposed to observe the Baily's beads, produced by the light from photosphere through the lunar valleys. In grazing eclipses their number N can be high, providing N determinations of photosphere's circle. But it is not their positions to be directly measured, since it should not be possible to do it better than 1 arcsec: it is the timing of appearing or disappearing of the Baily's beads. The actual radius of the Sun minimizes the scatter between calculated [e.g. with Occult 4 freeware program of D. Herald] and observed phenomenon, once considered also the opposite shadow's border. 
To reduce the effects of the uncertainties on lunar profile and lunoid corrections the eclipses data were preferably analyzed after a Saros (18 years and 11 days) which is a multiple of libration cycle, in order to observe the same Baily beads produced by the same lunar valleys. Another solution proposed by D. Dunham was to select only the polar beads, since the difference in latitude's libration among two eclipses is rather small and the same valleys produce the same beads at each eclipse\cite{Sigi09b}.    
The presence of the {\it Kiselevka valley} discovered during the total eclipse of 2008 in a place where the Watts atlas of the lunar profiles did not show any walley was published in 2009\cite{Sigi09c}. This is an example of a structure never identified in previous eclipse, nor in Watts profiles, therefore the method of polar beads still has some uncertainties.
The Kaguya's profile (published on Nov. 2009), by the name of the recent Japanese lunar mission, gives a profile which is expected to be more preciser than the Watts profile, even if the angular sampling is limited to a point each $\sim 16$ arcsec; its accuracy in height is $\delta h\pm 1$m.
A new era in the eclipse methods is started after Kaguya, but the solar limb definition in eclipses video is a new open problem, as we can see in the following paragraph.
Carles Schnabel\cite{Schnabel} and other observers since 2005 claimed out the visibility of chromosphere during annular eclipses, and after other observations of a thin region above the photosphere with telescopes ranging from 4.5" to 8" with neutral density filters (transmittance $\sim 10^{-4}$), the definition of a bead disappearance or appearance seems to be in need of revision.
Another example is the two observations of 2008 total eclipse made by Richard Nugent and Chuck Herold: the latter was more inside the umbral limit but observed with a 5" while the former observer used a 3" and did not see the light of the last thin layer above the photosphere. 

\begin{figure}
\centering
\includegraphics[width=12cm]{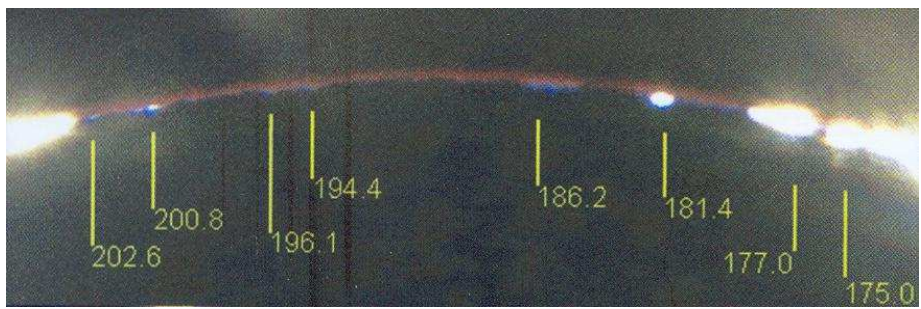}
\caption{Annular eclipse of 2005, photo by Carles Schnabel through a 4.5" Newtonian telescope. The numbers under the beads are the Watts angles with a corresponding lunar valley.
The whitish layer appears over 186.2; a small photospheric bead is on 200.8 another on 181.4; all the profile from 202 to 177 is reddish.
I my opinion this is a chromatism of the instrument acting on the white light coming from the layer immediately above the photosphere and under the chromosphere, enough luminous to be detected over the background due to the annularity of the eclipse. This chromatism appears also on the 
brilliant cusps at angles $\ge$203 and $\le$177. In this black and white edition of the image only a thin layer over the lunar profile is visible, as in the video of Chuck Herold published on youtube after 2008 and 2009 total eclipses.}
\end{figure}

\section{Flash spectrum and limb definition}
During an eclipse, the flash spectrum is the spectrum captured at the instants of beginning and end of totality.
W. Campbell when directed the Lick Observatory, steered some eclipse observational champaigns: the {\it Crocker's eclipses} from the person who financed them.
One of his experiments consisted in photographing over a moving plate the spectrum of the Sun through a slit perpendicular to the direction of motion of the Moon over the Sun. This experiment produced magnificent spectra, called spectrum flash.
The exposure of such images started 10 s before and end 10 s after the start of totality.
An image of this spectrum flash is published in the Lick Observatory studies of 1931\cite{Menzel} and here reprinted, see also\cite{Pannekoek,Campbell}.
  
\begin{figure}
\centering
\includegraphics[height=9cm]{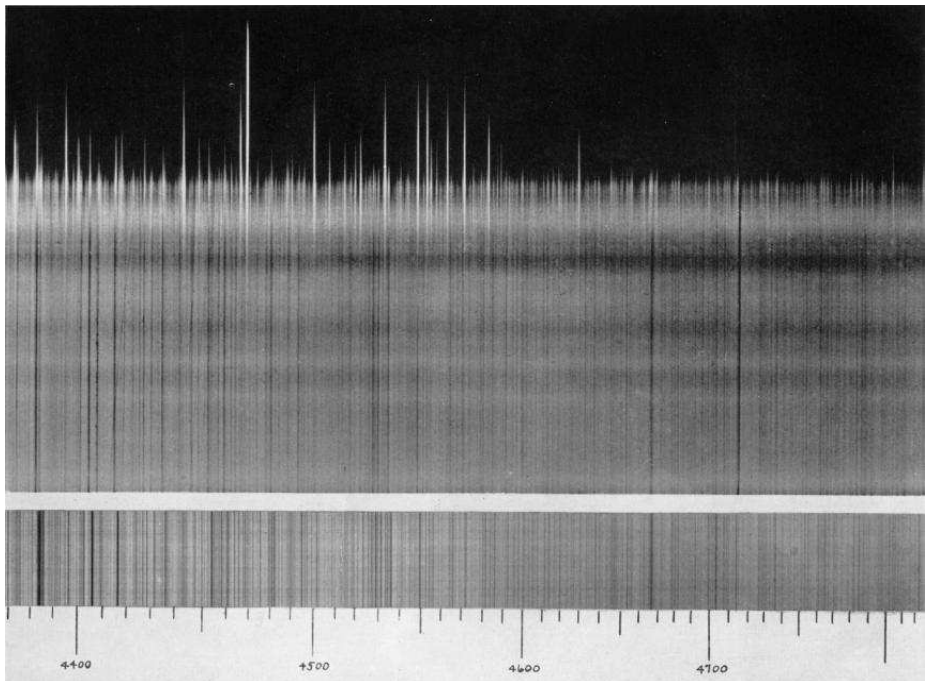}
\caption{Flash spectrum of 1905 total eclipse. The width of the slit was $w\sim 1.3$mm and the plate moved of $\sim w$ each second. The lower part of the figure is the spectrum before the totality, where the continuous of the photosphere and a lot of emitting lines. In the upper part there are only the lines of chromosphere; their blend is $10^{-4}$ times the continuum. After the beginnig of totaly only the continuum of the corona can be detected. The blend of the small emitting lines is perceived as white light and it is $\sim 10^{-3}$ times the intensity of the photosphere's continuum.}
\end{figure}

\section{Solar limb definition for drif-scan transits}
The Sun is a self gravitating gaseous structure, and its limit is not sharply defined, nevertheless the variation of the density with the height is exponential, and in the wavelengths of visible light the surface of unitary optical depth $\tau$=1 can be considered sharp with respect to the dark sky of the background.
The solar limb darkening function LDF, moreover, describes a decrease of the luminosity down to the $16\%$ of the value attained at the center of the disk. The combination of the LDF with the Point Spread Function PSF of the telescope pours photons out of the geometrical limb. The most suitable definition of solar limb has been chosen as the maximum of the derivative of the luminosity along a radius. This maximum can be detected by derivation of the Fourier anti-transform of the observational data, this method has been considered stable with respect to the seeing effects\cite{Hill}. Nowadays the influence of seeing on limb detection is being considered below the arcsecond level\cite{Irbah03,Sigi10}. How the Flash Spectrum Region and the atmospheric halos can affect this definition? It seems that the FSR and halo effects are negligible when the photosphere is visible, while during the very last phases of total eclipses FSR become important.
 
\section{Flash Spectrum Region and Baily's beads}
In fig. 7 is represented a scheme of the last solar features visibles during a total eclipse. The Baily's bead is a small sector of photosphere, already darkened to $16\%$ of its central luminosity\cite{Neckel}, and it is surrounded by two layers much fainter the Flash Spectrum Region FSR and the chromosphere.
The width of this FSR is within an arcsecond, while the chromosphere goes up to three arcseconds.
The intensity of the FSR integrated over the area can be larger than the residual luminosity of the bead, and this happen frequently with big telescopes observing grazing eclipses (examples of C. Schnabel and C. Herold). This pheonomenon could have been determinant in the hybrid (annular-total) eclipse observed by Clavius in 1567\cite{Clavio}, to explane the observed annularity in contrast with the calculated totality by more than 4 arcseconds.
Another problem for the dis/appearing bead can be the scintillation\cite{SigiNugentDangl}.
\begin{figure}
\centering
\includegraphics[width=7cm]{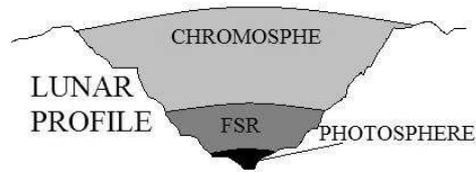}
\caption{The intensity of the chromosphere, coloured by particular emission lines, is $10^{-4}$ times the intensity of the photosphere (central value). The Flash Spectrum Region FSR is $10^{-3}$ times the intensity of the photosphere. There is a confusion limit for which the intensity of the bead (photosphere)
equals the light from all FSR visible. It is necessary to take this fact into account when the dis/appearance timing of a bead is studied with a few arcseconds of space resolution imaging. Opportune models and scale heights have to be set up. Data from 2010 total eclipses and experiments carried on by Serge Koutchmy and Cyril Bazin will help us in this respect.}
\end{figure}

\section{Conclusions}
The next total eclipses will provide us important data on the FSR, and the dataset of Baily beads\cite{Sigi09c} will be revisited, in order to better understand and treat ancient eclipses and to recover the past behaviour of solar diameter.
PICARD satellite mission, expected to start in June 2010\cite{picard}, is expected to provide a milliarcsecond precision on the solar diameter measurements.

\subsubsection*{Acknowledgments.} It is a pleasure to record my thanks to Dr. Michele Bianda, director of the IRSOL observatory; to Ing. Cyril Bazin who provided me with the observations at the Carte du Ciel in Paris and to Prof. Serge Koutchmy who set up this experiment at the Observatoire de Paris and who invited me there from April 30 to May 14, 2010, and to Dr. Patrick Rocher of IMCCE for a fruitful discussion on ephemerides.


\begin{thebibliography}{4}

\bibitem{Auwers}Auwers, A., Astron. Nachr. 128, 361 (1891)
\bibitem{Goettingen}Schur, V. and L. Ambronn, Astron. Mitt. der K. Sternwarte zu Goettingen 7, 17 (1905)
\bibitem{Sigi02}Sigismondi, C., Am. J. Phys.70,1157 (2002)
\bibitem{Sigi06}Sigismondi, C., Proc. 233 IAU Symposium, 521 V. Bothmer and A. Hady eds., Cambridge Un. Press (2006)
\bibitem{Gething}Gething, P., Mon. Not. R. Astron. Soc. 115, 558 (1955)
\bibitem{Doraysol}Definition et Observation du RAYon SOLaire \url{http://www.oca.eu/gemini/equipes/ams/doraysol/index.html}
\bibitem{SolObs} Sigismondi, C., Overview on Solar Diameter Measurements, Solar Observer 1, 17-22 (2010) \url{http://solar-observer.com}
\bibitem{Sigi09}Sigismondi, C., Sci. China G 52, 1773 (2009)
\bibitem{SigiNugentDangl}Sigismondi, C., R. Nugent and G. Dangl, Proc. III Stueckelberg Meeting, Pescara (2010)
\bibitem{Fivian}Fivian, M. D., et al. Science 322, 560 (2008)
\bibitem{Sigi05}Sigismondi, C., Proc. X Marcel Grossmann Meeting on GR, Astro-ph0501291 (2005)
\bibitem{Sigi05b}Sigismondi C. and P. Oliva, N. Cim. B 120, 1181 (2005)
\bibitem{Dicke}Dicke, R. H., Astrophys. J. 159, 1 (1970)
\bibitem{Kuhn04}Kuhn, J. R.,  et al., Astrophys. J. 613, 1241 (2004)
\bibitem{Kuhn10}Kuhn, J. R.,  et al., Astrophys. J. in press (2010)
\bibitem{Eddy}Eddy, J.A. and A. A. Boorrnazian, Bull. Am. Astron. Soc. 11, 437 (1979)
\bibitem{Sigi08}Sigismondi, C., M. Bianda and J. Arnaud,  AIP Conference Proceedings, 1059, 189 (2008)
\bibitem{Watts}Watts , C. B., The Marginal Zones of the Moon, USNO (1963)
\bibitem{kaguya}Kaguya-Selene Japanese Satellite Mission \url{http://www.selene.jaxa.jp/index_e.htm}
\bibitem{MA}Morrison, L. V. and G. M. Appleby, Analysis of Lunar Occultations - Part Three - Systematic Corrections to Watts' Limb Profiles for the Moon, MNRAS 196, 1013--1020 (1981)
\bibitem{soma}Soma, M., Limb profiles of the Moon obtained from grazing occultation observations, Publ. Natl. Astron. Japan, 5, 99--119 (1999).
\bibitem{Paddock}Paddock, G. F., The Chromospheric Spectrum as Observed with an Objective Prism at the Eclipse of January 24, 1925, Astrophys. J. 66, 1--12 (1927)
\bibitem{Sigi09b}Kilcik, A., C. Sigismondi, J.P. Rozelot and K. Guhl, So. Phys. 257, 237 (2009)
\bibitem{Sigi09c}Sigismondi, C., D. Dunham et al., So. Phys. 258, 191 (2009)
\bibitem{Schnabel}Schnabel, C., Observacion de Eclipses de Sol en los Limites de la Centralidad, Trabajos de investigacion II, 46-69 \url{http://www.astrosabadell.cat}(2009)
\bibitem{Menzel}Menzel, D. H., A study of the flash spectrum, PAAS 6 145 (1931)
\bibitem{Campbell} Campbell, W. W. and C. D. Perrine, The Lick Observatory Crocker Eclipse Expedition to Spain, Proc. Astron. Soc. Pacific 18, 13--36 (1906)
\bibitem{Pannekoek}Pannekoek, A. and  M. G. J. Minnaert, Results of observations of the total solar eclipse of June 29, 1927. I. Photometry of the flash  spectrum, Verhandelingen der Koninklijke Akademie van Wetenschappen te Amsterdam, XIII 1--106 (1928)
\bibitem{Hill}Hill,H.A., R. T. Stebbins and J. R. Oleson, Astrophys. J. 200, 484 (1975)
\bibitem{Irbah03}Irbah A., et al., in Lecture Notes in Physics, 599, 159 (2003)
\bibitem{Sigi10} Sigismondi, C., Daytime seeing and solar limb positions: this volume (2010)
\bibitem{Neckel}Neckel, H. and K. Slab, So. Phys. 153, 91 (1994)
\bibitem{Clavio}Stephenson, F. R., J. E. Jones, and L.V. Morrison, Astron. and Astrophys. 322, 347 (1997)
\bibitem{picard}PICARD, CNES Satellite Mission \url{http://smsc.cnes.fr/PICARD/}

\end{thebibliography}
\end{document}